# A Modern Laboratory XAFS Cookbook

**GT Seidler**[(\*)]**, DR Mortensen, AS Ditter, NA Ball, AJ Remesnik**

Department of Physics, University of Washington
Seattle, Washington, United States 98195-1560

[(\*)]E-mail: seidler@uw.edu

**Abstract**. We have recently demonstrated a very favorable, inexpensive modernization of lab-based x-ray absorption fine structure (XAFS) and high-resolution x-ray emission spectroscopy (XES) using only commercially-available optics and x-ray tube sources. Here, we survey several proven instrument designs that can be readily implemented in any laboratory setting to achieve synchrotron-quality XAFS and XES for many systems in the 5 keV to 10 keV energy range. These approaches are based on our immediate experience with the development of: (1) an inexpensive, low-powered monochromator capable of performing either XAFS or XES, (2) a mid-scale XAFS user facility having $10^6$/sec flux with sub-eV bandwidth on each of two independent beamlines, and (3) multiple XES spectrometers having $3^{rd}$-generation synchrotron performance for battery and actinide research.

## 1. Introduction

The advent of synchrotron light sources has led to the steady growth and impressive scientific reach of XAFS. While the synchrotron light sources provide a gold-standard of technical capability and user support, the fact remains that the nearly total localization of XAFS to the synchrotron facilities is an anomaly in current scientific practice. By contrast, none of laser spectroscopies, NMR, electrical transport, or x-ray diffraction would be so broadly important across the sciences, engineering, and technology if students and future users could train and use only a few dozen, heavily subscribed facilities worldwide. For most techniques, there is a range of instrumentation spanning the reciprocal continuum of access to cost and capability. Inexpensive, benchtop systems are widely distributed and can be used for routine studies, education, and initial sample characterization, while only the most advanced systems exist as limited-access world-class shared-user facilities or as unique, expensive research instruments.

Building on the above observations,[1] we have argued that the access limitations inherent to synchrotron facilities adversely impacting active XAFS research programs and the education of the next generation of XAFS users, but also preemptively exclude a large body of scientific studies or industrial research and development. This includes: work by scientists in developing countries who lack the travel funding needed for frequent synchrotron visits, studies requiring complex *in situ* apparatus that are not easily transported to the light sources, and system-specific considerations that inhibit sample transport to the light source, *e.g.*, extreme air sensitivity or radiological safety considerations. The same concerns have existed from the earliest days of XAFS synchrotron beamline development.[2]

For these reasons we have begun a long-term research program [1] aimed at rejuvenating laboratory-based XAFS and also enabling laboratory-based high resolution x-ray emission spectroscopy (XES); note that we discuss XES in another entry to the present proceedings.[3] This paper continues as follows. In section 2, we review some basic issues involved in the operation of Rowland circle monochromators. In section 3, we discuss the design and performance of our low-powered XANES and XES system that has now been operating for over a year. In section 4, we present an overview of the design of a high-powered two-beamline user facility that is now well into commissioning at the University of Washington. In section 5, we briefly discuss a new single-motor drive system and some future directions. Finally, in section 6 we summarize and conclude.

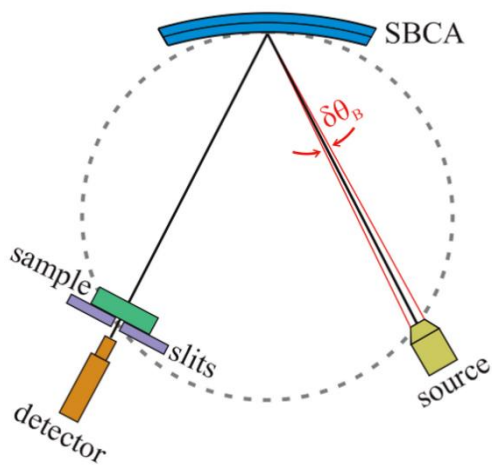

**Figure 1**: Schematic of a Rowland circle monochromator using a spherically bent crystal analyzer (SBCA). Polychromatic bremsstrahlung from a conventional x-ray tube is incident on the SBCA. In the plane of the figure, excluding the small Johann error from the optic, the radiation is refocused to the symmetrically-located exit point on the Rowland circle but is also wavelength-selected via Bragg scattering. Energy scanning uses symmetric motion of the source and detector on the Rowland geometry. For transmission-mode XAFS, as shown in the figure, it is convenient to place the sample in front of the exit aperture (slits) placed on the Rowland circle in front of the detector.

## 2. The Rowland Circle Monochromator

The problem of point-to-point monochromatization of hard x-rays is widespread, and decades of instrument development across numerous platforms support the excellent function and versatility of the Rowland circle configuration, as depicted in Fig. 1. In such a system the source, detector and optic are all placed as shown on a circle whose diameter must match the radius of curvature of the optic. Cylindrical optics were used in the earlier generation of laboratory XAFS systems and still see some use in high-resolution applications,[4] but the overwhelming majority of present-day, high-resolution Rowland circle spectrometers are instead based on spherically-bent crystal analyzers (SBCA).

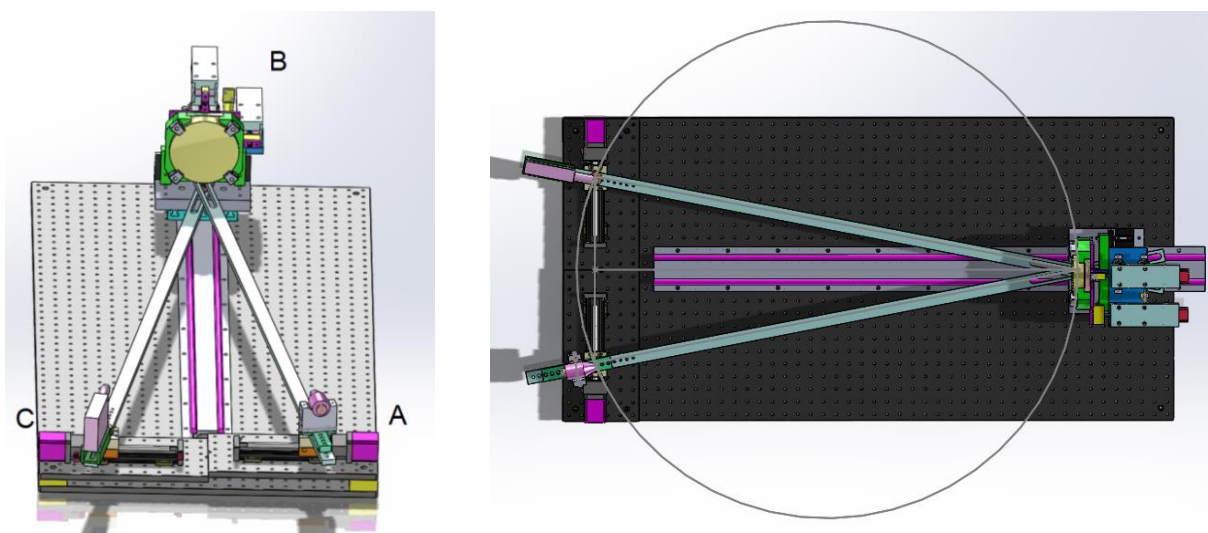

**Figure 2**: *Left*: A perspective rendering of the low-powered lab XANES system. (A) The source positioner with the x-ray tube mounted on a V-block on the source steering bar; (B) SBCA mounted on a 2-axis tilt stage which is in turn mounted on a passive linear slide; (C) The detector positioner with the detector mounted onto a hold on the detector steering bar. For clarity of presentation the helium flight path and radiation enclosure have been suppressed. *Right*: A top-view rendering of the same instrument with the 1-m diameter Rowland circle superimposed.

There are many ways to implement the scanning motion on the Rowland circle in the laboratory frame, with the main consideration being whether some benefit accrues from having one of the major components held stationary in the laboratory frame. For example, the low-powered instrument described in section 3 holds no component fixed in the lab frame while the XANES user-facility and high-powered XES systems in sections 4 and 5 keep the (fairly heavy) source fixed. No matter the choice of drive mechanism, when working not too far from backscatter the energy resolution is still dominated by the source size in the plane of the Rowland circle. Consequently it is the proliferation of available SBCA crystal orientations that helps enable the present effort; SBCA can be fabricated or purchased having Bragg angles of 70-85 degrees for almost any absorption edge in the hard x-ray range.[5] This allows ~1-eV energy resolution in the hard x-ray range when using ~0.5 – 1 mm source spot sizes, greatly expanding the useful selection of inexpensive commercial x-ray sources.

## 3. Low-powered Monochromator/Spectrometer for XAFS and XES

Our entry into laboratory-based spectroscopy was somewhat inadvertent. We sought to have in-house capability to test x-ray optics, either in support of mass-scale manufacture of optics for a next-generation multi-analyzer system for inelastic x-ray scattering [6] or of a project on time-resolved XAFS and XES of luminescent lanthanide compounds [7]. A broad survey of commercial x-ray sources found a family of inexpensive low-powered air-cooled x-ray sources that appealed on the basis of convenience and source spot size. To our surprise, an order of magnitude calculation of flux for a 1-m Rowland circle predicted excellent performance for XES and useful capability for transmission-mode x-ray absorption near edge structure (XANES). The decision was quickly made to develop such an instrument and in the subsequent 60 days all components were acquired and the instrument was assembled, computer-interfaced, and safety certified. The construction of a laboratory-based, high-resolution Rowland-circle monochromator is not a major project in cost, time, or manpower.

In figure 2 we present CAD renderings of this instrument, somewhat modified from the original report.[1] Relative to the earlier iteration, the present system instead mounts the SBCA 2-axis tilt unit onto a passive (non-motorized) linear slide. Positioning of the SBCA along this slide is then set by a system of three conjoined steel wires (not shown), each of length 0.5 m, anchored to the pins under the source, detector, and SBCA locations. The wire triad's raw geometrical constraint forces the three components to be on a 1-m diameter circle, and the alignment of the system establishes the requisite symmetry on the circle. Energy scanning involves synchronous scanning of the source and detector location and is accompanied by a translation of the Rowland circle in the lab frame.

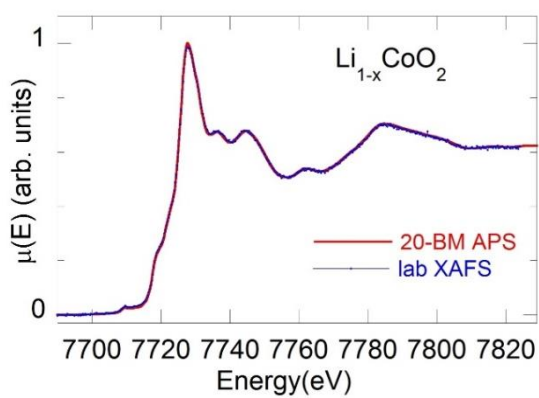

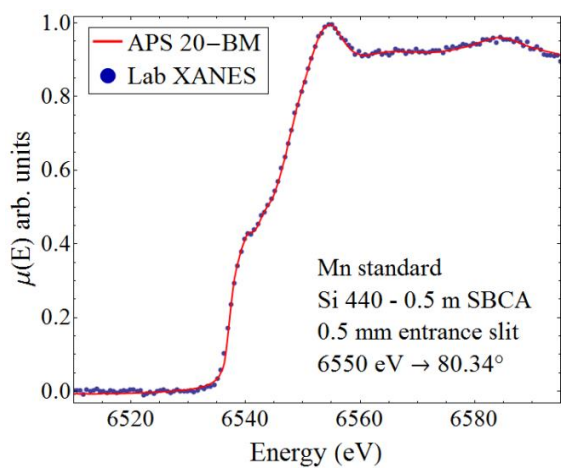

**Figure 3**: Transmission-mode XANES for an *ex situ* battery electrode material. The data indicated by the red curve was taken at beamline 20-BM of the Advanced Photon Source while the blue data (small dots) was taken with the lab XANES instrument. The optic was Ge (444) with a 1-m radius of curvature. No source slit was used.

**Figure 4**: Transmission-mode XANES for an Mn reference standard. The data indicated by the red curve was taken at beamline 20-BM of the Advanced Photon Source while the blue data (small dots) was taken with the lab XANES instrument. The optic was Si (440) with a 0.5-m radius of curvature and a 0.5-mm wide source slit was used.

As shown in figures 3 and 4, the performance of this system for transmission-mode XANES of concentrated samples is very favorable. Note that the results in figure 4 were taken with one of the 10-cm wafer diameter, 0.5-m radius of curvature SBCA (XRS Tech) that are only now becoming available; the obvious modifications where made to achieve the tighter Rowland circle. If such optics become more common, the overall instrument footprint could be correspondingly decreased for a more truly "benchtop" system, in addition to having a larger collection solid-angle.

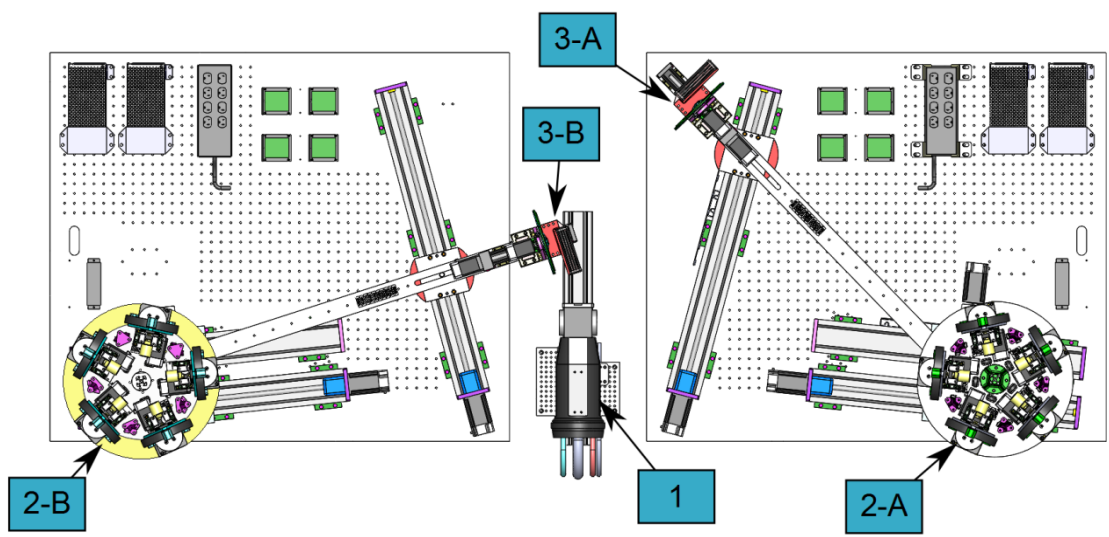

**Figure 5**: A top-view rendering of the CEI-XANES instrument. (1) X-ray tube housing with shutters in both horizontal directions, 6-degree take-off angle. (2-A) and (2-B) analyser optic turret. (3-A) and (3-B) detector sub-assemblies including sample turret. The radiation enclosure, helium flight path, and additional support structure have been supressed for clarity of presentation.

## 4. The University of Washington Clean Energy Institute XANES User Facility

XAFS has proven to be a powerful tool in the world-wide effort to improve electrical energy storage for vehicle electrification and grid-level storage. This situation presents an important example of how laboratory-based XAFS can have high scientific and social impact. The ongoing research work has the common character of materials chemistry: rapid iteration or hypothesis, synthesis, and evaluation is needed, but such rapid turn-around is typically problematic, albeit not impossible, at the synchrotron light sources. However, the best research battery cell, the so-called 'pouch cell', can have outstanding agreement with the technical demands of transmission-mode XAFS studies and consequently with lab-based XAFS applications, thus meeting the demand for rapid feedback and ultra-high access.

It is this rationale that motivated the ongoing development of a two-beamline lab XAFS user facility at the University of Washington, funded by the UW Clean Energy Institute. This instrument, referred to as CEI-XANES although its available energy scanning range will often extend into the EXAFS regime, is now almost fully assembled and one of the two monochromators is very near to completion of commissioning. CAD renderings of this instrument are presented in figures 5 and 6. For the purpose of the present manuscript, i.e., the discussion of different designs of lab XAFS instruments, the most important technical point is that this system is almost entirely 'modern', in that rather than using mechanical linkages to enforce the Rowland circle constraints upon energy scanning the CEI-XANES instrument must instead synchronize multiple non-symmetry-related motor drives. This design choice followed from the need to have an SBCA turret to enable easy selection of energy range and thus simplify user operation and staff support. In this case, it is difficult, e.g., to passively track the Bragg angle of the SBCA to half-angle the detector arm motion.

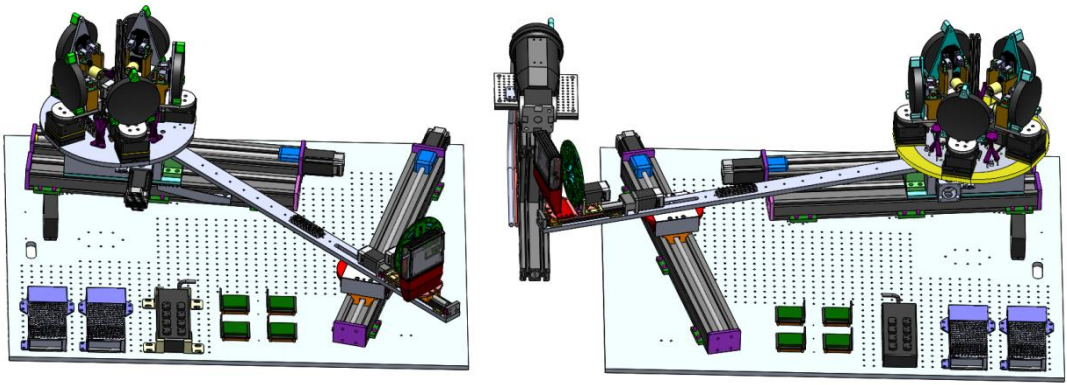

**Figure 6**: CEI perspective: A perspective rendering of the CEI-XANES instrument. The two monochromators operate fully independently. The radiation enclosure, helium flight path, and additional support structure have been supressed for clarity of presentation.

CEI-XANES commissioning has been more challenging than that for the simpler instrument of the preceding section. Considerable effort must be placed on coordinating the motors for a systematic-free calibration and also the demands on smoothness of motion of the Bragg-angle stage under each SBCA are more strict than originally anticipated. At the time of submission of this proceeding, these challenges are almost completely overcome. Good motor tracking and the flux milestone of $10^6$/sec has been reached in the 5-10 keV range and *in situ* studies of pouch cell batteries will being shortly. A more complete report on this instrument will follow, but we find the effort already to be a significant success.

## 5. Revisiting the Linear XAFS Spectrometer

Given the above *caveats* on the full nominal modernization to motorization and computer control over multiple degrees of freedom to drive energy scanning on the Rowland circle for systems where the x-ray source must be fixed in the lab frame, it is interesting to revisit the classic 'linear spectrometer' design from the 1970's and 1980's,[8-10] updated only to operate closer to backscatter and use SBCA rather than the older cylindrical optics, thus allowing high energy resolution at all absorption edges. The key point is that energy scanning of these instruments can be achieved by a single motorized lead-screw drive with smooth tracking of the SBCA orientation driven by high-quality mechanical linkage. Three such systems are now under construction in our group; one for in-house use, one that will become an XES instrument for the study of batteries in the user facility of the Joint Center for Energy Storage Research at Argonne National Labs, and one that is destined for XES of *f*-electron materials at Los Alamos National Labs. A CAD rendering of these systems in the XES configuration is presented in Figure 7. Switching to a generic line-focus few-kW tube in line-of-sight to the optic, in analogy to the CEI-XANES instrument, converts this XES spectrometer to an XAFS monochromator. Full details of this design will be published after commissioning in the autumn of 2015.

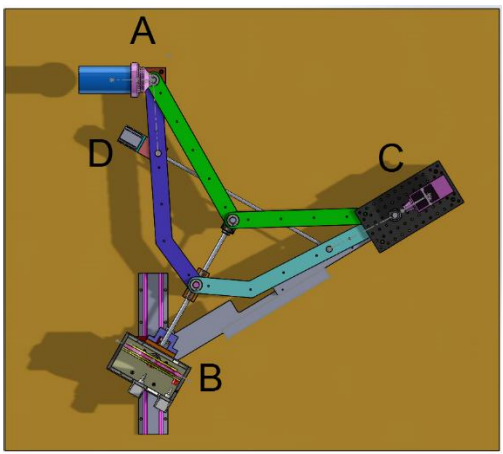

**Figure 7**: A top-view CAD rendering of a 'linear spectrometer' design under construction. (A) X-ray source and sample for XES; (B) SBCA 2-axis tilt on passive linear slide; (C) Detector platform; (D) Stepper motor drive. The stepper-driven leadscrew pushes the detector and source arms apart, with the shown mechanical couplings enforcing both placement and $\Theta$-$2\Theta$ symmetry on the Rowland circle.

Before concluding, in the context of our 'modern cookbook' one important component has not been discussed: the ready availability of high spatial resolution x-ray cameras. As investigated by other speakers at this conference, this allows numerous wavelength-dispersive designs having their own unique performance advantages and disadvantages. We particularly endorse the use of such systems at energies lower (few keV) and much higher (15-20 keV) than considered here. In the former case, this approach has led to outstanding performance for XES [4] and the switch to a bremsstrahlung source at the sample location would convert such an instrument to an interesting XAFS system. For the latter case, early reports of high-quality lab-based EXAFS at energies needed for actinide $L_3$-edge studies would benefit from this approach and also a possible change of the Laue optic to a curved geometry.[11]

## 6. Conclusions

We have reported some of our experience in the development of point-focusing, Rowland-circle monochromators for laboratory-based XAFS and XES instruments. Each of the source, optic, and detector characteristics have diversified in economically- and technically-favorable ways in the few decades since lab-based XAFS fell out of favor. We propose that the high demand for even 'routine', transmission-mode XAFS of concentrated samples supports the rebirth and rapid spread of lab-based systems and user facilities. Our simple, low-powered system is particularly favorable for this purpose.

## 7. Acknowledgements

This work has been supported by: (1) the U.S. Department of Energy, Basic Energy Sciences under Grant No. DE-FG02-09ER16106 and also by Fusion Energy Sciences and the National Nuclear Security Administration thought Grant No. DE-SC0008580; (2) the State of Washington through the University of Washington Clean Energy Institute; (3) The Joint Center for Energy Storage Research (JCESR), an Energy Innovation Hub funded by the U.S. Department of Energy, Office of Science, Basic Energy Science; and (4) The U.S. Department of Defense through a subcontract from Los Alamos National Laboratories. APS Sector 20 operations are supported by the US Department of Energy and the Canadian Light Source, with additional support from the University of Washington.